\documentclass[10pt,prb,showpacs,twocolumn,amsmath,amssymb,floatfix]{revtex4-1} 
\usepackage{graphicx}
\usepackage{dcolumn}
\usepackage{bm}
\usepackage{hyperref}
\usepackage{longtable}
\begin{document}
\author{Basudeb Mandal}
\author{Hirak Kumar Chandra}
\thanks{BM and HKC have contributed equally to this work.}
\author{Priya Mahadevan}
\affiliation{Department of Condensed Matter Physics and Material Sciences, S.N. Bose National Centre for Basic Sciences, Block- JD, Sector-III, Salt Lake, Kolkata- 700098, India}
\title {Quantum confinement : A route to enhance the Curie temperature of Mn doped GaAs}

\begin{abstract}
The electronic structure of Mn doped GaAs and GaN have been examined within a multiband Hubbard model.
By virtue of the positioning of the Mn $d$ states,
Mn doped GaAs is found to belong to the $p$-$d$ metal regime of the Zaanen-Sawatzky-Allen phase diagram
and its variants while Mn doping in GaN belongs to the covalent
insulator regime. Their location in the phase diagram also determines
how they would behave under quantum confinement which would increase the
charge transfer energy. The ferromagnetic stability of
Mn doped GaAs, we find, increases with confinement therefore providing
a route to higher ferromagnetic transition temperatures.
\end{abstract}
\pacs{}

\maketitle

While early research on dilute magnetic semiconductors focussed on II-VI 
semiconductors\cite{Furdyna-1988,Corbett-1983}, the possibility 
of using low temperature molecular
beam epitaxy for the growth shifted the focus onto 
III-V semiconductors\cite{Ohno-1996, samarth-2002,nature-2005, rmp-2014, freeland-2003}.
Additionally the growth technique allowed the introduction of transition 
metal atoms far beyond their equilibrium solubility. The quest for
higher concentrations of the transition metal in the semiconductor was
driven by considerations of increasing the magnetic ordering temperature.
However initially various material issues plagued the discovery of new dilute 
magnetic semiconductors and one was never sure if the magnetism was
intrinsic or due to impurity phases. With recent developments in both
the growth as well as the characterization of dilute magnetic
semiconductors, puzzling observations are emerging which don't seem
to be explainable within the current models used to explain 
magnetism. Considering the well studied example of GaAs doped
with Mn, which is believed to represent a system where the magnetism is
intrinsic, the highest Curie temperature has been found to 
be around 185 K\cite{Novak-2008}.
However, recent photoemission experiments found spin polarized bands even
at room temperature\cite{arxiv-2014} and this can't be reconciled with the existing models
for ferromagnetism. 

Further the same transition metal atom doped in different semiconductors 
has led to varying behavior \cite{APL-2001,gamnn}. This was initially explained within the kinetic
exchange model in which each transition metal atom was approximated by a spin.
This spin interacted with the valence band of the host semiconductor,
resulting in a spin polarization of the carriers which
mediated the magnetism \cite{Dietl-1997}. 
Later models questioned this hypothesis
and put forth the picture of an impurity band emerging from the
interaction of the transition metal with the host 
semiconductor\cite{Mahadevan-2004,APL,RKKY}. 
However this description has remained at a qualitative level\cite{APL,RKKY}. 
The picture of the impurity band emerging in GaAs doped with Mn 
has been supported by recent experiments \cite{tanaka,dobrowolska,fujii}. 
There have 
also been improved approaches to examine the electronic structure\cite{fujii,dmft}, 
however, there is still no consensus on the mechanism of magnetism.

Bulk transition metal compounds
have been studied for a long time and their electronic structure is
well established within the framework of the Zaanen-Sawatzky-Allen (ZSA)
phase diagram\cite{ZSA-1985,Nimkar-1993,DD-pramana}. A similar framework
should be possible for dilute magnetic semiconductors which we examine
in the present work. We consider a multiband Hubbard model to describe
the electronic structure of the dilute magnetic semiconductor with
Coulomb interactions included on the transition metal site.
Parameters appropriate for Mn doped GaAs place it in the regime of a
$p$-$d$ metal of the ZSA phase diagram, thereby explaining why correlation 
effects don't drive it insulating. Quantum confinement effects can
be used to tune the charge transfer energy $\Delta$, driving a change 
in the character of the hole state. We show that this 
also serves as a parameter to change the Curie temperature, with the 
largest ferromagnetic stability being in the vicinity of 
$\Delta_{eff}$ equal to zero. $\Delta_{eff}$ is the charge transfer 
energy defined 
between the Mn $d$ states and the dangling bond states with $t_{2}$ symmetry.
Confinement arising from the presence of the
surface could lead to the same effect of 
enhancing the Curie temperature, therefore
explaining the experimental observation of spin polarized bands 
even at room temperature\cite{arxiv-2014}. Mn doped GaN 
is found to be a covalent insulator within the framework of 
our calculations, thereby explaining the different behavior found for
Mn doping in GaAs and GaN.

In order to discuss various aspects of the electronic structure 
of Mn doped GaAs, a multiband
Hubbard model is set up to solve the electronic structure. In 
this model, $d$ states are included on the Mn atom, $s$,$p$
states are included on the Ga atoms while $s$,$p$,$d$ states are included on the As atoms. 
Hopping is allowed between nearest neighbor Mn and As atoms, 
between Ga and As atoms as well as between nearest neighbor Ga-Ga and 
As-As atoms. The onsite energies
as well as the hopping interaction strengths are determined by fitting 
the ab-initio band structure \cite{suppl-info} of nonmagnetic 25\%
Mn doped GaAs calculated within VASP\cite{Joubert,vasp,vasp1} to a tight-binding model\cite{PRB-TB}. 
Onsite Coulomb interactions between the $d$ orbitals on Mn are 
parametrised in terms of the Slater-Condon
integrals $F^0$, $F^2$ and $F^4$. While $F^2$ and $F^4$ are fixed at 80$\%$ of their atomic Hartree-Fock values, $F^0$
is fixed so that the multiplet averaged $U$ is at a pre-determined value. In the rest of the discussion, we use only the
multiplet averaged $U$. 
A similar tight binding fitting of the ab-initio band structure for nonmagnetic 25 $\%$ Mn doped GaN is carried
out to determine the one electron part of the multiband Hubbard Hamiltonian for Mn doped GaN which is then solved.
A mean field decoupling scheme
has been used for the four fermion terms in the Hamiltonian which is solved self-consistently for the order parameters as discussed
earlier\cite{La2NiO4-1997,Nimkar-1993} over a 4x4x4 k-points grid for the 64 atoms supercell. In order to explore magnetism,
a spin spiral implementation is considered which
uses the generalised Bloch's theorem so that the same unit cell could be used for different magnetic configurations
characterised by the wave vector $q$\cite{eschrig-1998}.

Early work on transition metal compounds established the important 
role that correlations played in driving the system insulating\cite{minami}. 
As the nature of the ground state was
largely determined by electron-electron interactions at the 
transition metal site, it was a surprise when it was found that on 
changing the anion in a set of late 3$d$ transition metal compounds, 
one had large changes in the bandgaps, with even metallic members 
being found\cite{minami}. This established that in addition to the onsite Coulomb 
interactions ($U$), there was another scale in the problem, which 
was the cost of charge 
transfer ($\Delta$) between the anion $p$ states and the transition 
metal $d$ states. This was brought out by Zaanen, Sawatzky and Allen\cite{ZSA-1985}
in their seminal phase diagram which now forms the basis for classifying 
the electronic structure of transition metal compounds. 
We now consider examples of two well-studied semiconductors - GaAs:Mn and GaN:Mn and
examine in which regime they lie in the ZSA phase diagram.

The band dispersions for nonmagnetic 25$\%$ Mn doped GaAs are given in Fig. 1. The 
ab-initio band dispersions are given by the dashed blue lines, while the best
tight-binding fit are given by the solid red lines. The fitting procedure
involves an optimization of the best fit band structure along various
symmetry directions by a least square error minimization process\cite{PRB-TB}. The
bare charge transfer energy between the Mn $d$ states and the As $p$ states 
estimated from the fitting is found to be
0.53~eV. Other parameters extracted from the fitting are given in the
Supplementary Information. We then use these parameters as the tight
binding part of a multiband Hubbard Hamiltonian and calculate the
electronic structure of Mn doped GaAs at a doping percentage of 3.125$\%$
which is within the range of concentrations probed in experiments. The calculated partial
density of states is shown in Fig.~2 where the zero of the energy scale
is the Fermi energy. One finds that the up spin states with dominantly Mn
character lie deep inside the valence band with some weight at the
Fermi energy also. The As atoms which are the nearest neighbors of the Mn
atom are found to contribute primarily to the state at the Fermi level
while those atoms which are far away have a weak contribution. The
spin polarization  of the states localized on the nearest neighbors
of the Mn atom is large while it is weaker for the states associated
with the farther away As atoms. This is consistent with the impurity
model description\cite{Mahadevan-2004,RKKY,APL} introduced by 
Mahadevan and Zunger in which the 
electronic structure of Mn doped GaAs can be visualised as arising
from the interaction of the $d$ levels on Mn with the states
present prior to the introduction of the Mn atom at a Ga site (i.e
the dangling bond states associated with a Ga-vacancy). The
interaction is primarily between the levels with $t_2$ symmetry 
on Mn and the levels with the same symmetry on the dangling bonds. 
These dangling bond states are dominantly localized on the As atoms which
are the nearest neighbors of the Mn atom. Consequently one finds that the
states at the Fermi energy which are the antibonding states of these 
interactions are localized on these atoms.
A $U$ of 3~eV is used on Mn, though we have also 
increased the value from 3~eV to
4~eV to examine its effect on the electronic structure. The
calculated Mn $d$ partial density of states for $U$=4~eV is shown in the
inset of Fig. 2(a) and the system is still metallic.
Hence considering reasonable values of U on Mn does not drive Mn doped GaAs insulating.

The definition of $\Delta$ should be with respect to the energy of the
dangling bond states with $t_2$ symmetry and is referred to as 
$\Delta_{eff}$ in the subsequent discussion. As this is difficult to
determine precisely, we vary $\Delta$ and examine the character
of the hole state. When the two interacting levels are degenerate
($\Delta_{eff}$=0) one expects that the hole
has equal weight on Mn as well as the interacting As $p$
atoms. In Fig. 3 we have plotted the variation of the Mn $d$ partial
density of states as a function of $\Delta$. As $\Delta$ is 
increased, one finds an increase in the Mn $d$ contribution to the 
hole state. Tracking the Mn $d$ character of this state ($n_d$) (inset of Fig. 3),
one finds that between $\Delta$ of 2.6 and 2.7 eV one has a change over
with the hole becoming predominantly Mn $d$ like. This places $\Delta_{eff}$=0
near a $\Delta$ of 2.7 eV in contrast to the value of 0.53 eV found from the fitting. Hence Mn doped GaAs is in the negative $\Delta$
regime of the ZSA phase diagram. As it is metallic, we identify it as a 
$p$-$d$ metal. While the charge transfer energy is usually a fixed 
quantity for a system, here, $\Delta_{eff}$ is referenced with respect to
the dangling bond states. The dangling bond states follow the valence band maximum of the
host semiconductor. The latter can be tuned in a semiconductor by
quantum confinement\cite{Q-confinement}. The state corresponding to the
bulk case ($\Delta$=0.53~eV) may be described by the 
configuration $d^5\underline{L}$
where locally Mn is found to be $d^5$, and so is Mn$^{2+}$-like. In the
regime where $\Delta_{eff} \ge$ 0, Mn may be represented by the configuration $d^4$
and is therefore in the valence state Mn$^{3+}$. Hence one
has a valence transition with decrease in size of Mn doped GaAs. The
valence transition has been discussed earlier in the literature by
Sapra {\it et al.}\cite{sapra-2002} using a tight binding model. What we show is that
a metal-insulator transition accompanies this valence transition.

The immediate question which follows is how does the 
stability of the ferromagnetic state change
with quantum confinement. We examine this by considering an 
isolated Mn impurity in the 64 atom
unit cell (i.e, a doping percentage of 3.125 $\%$) and 
comparing the energies of the ferromagnetic
as well as the totally antiferromagnetic configuration given by $q$=0.5 0.5 0.5.
The Mn atoms are separated by 11.3 $\AA$. The interaction between them is
weak and so for small $\Delta$ values, one finds that the different 
magnetic solutions have comparable energies (Table I). For larger values we have
the system favoring an antiferromagnetic ground state. This is in
contrast with experiments which find ferromagnetism\cite{APL-2001}.
However Mn atoms show a tendency to cluster\cite{APL-cluster,raebiger3}
and the high magnetic ordering temperatures observed have been
associated with the presence of
these clusters\cite{akash,APL-2004,APL-2006}. 
We therefore went on to examine the 
variations in the ferromagnetic stability
by considering pairs of Mn atoms occupying FCC nearest neighbor 
positions as well as fourth
neighbor positions where the separations are 3.995 \AA{} and 7.99 \AA{} respectively. The difference
in energy between the ferromagnetic as well as antiferromagnetic
configuration are given in Table II. As $\Delta$ is increased, one finds
that the ferromagnetic stability increases till a $\Delta$ of 2.3~eV for pairs 
of Mn atoms at first neighbor positions and then it begins to decrease.
This can be traced to the fact that for a $\Delta$ of 2.7~eV we had
the Mn $t_{2g}$ and dangling bond states almost degenerate for 1 Mn.
The presence of the second Mn at the nearest neighbor position changes
some details of the $\Delta$ at which the two interacting levels are degenerate.
The result is that the ferromagnetism is stabilized by 
superexchange between the Mn atoms involving the intervening As atom,
explaining the enhanced ferromagnetic stability at $\Delta_{eff}$=0.
The calculated ferromagnetic stability at fourth neighbor also shows
a similar trend, being largest at $\Delta$=2.6~eV.

Mn doping in bulk GaAs has a ferromagnetic stability of 73 meV 
for nearest neighbor atoms. The increase
till  $\Delta_{eff} \sim $ 0 is approached reflects the 
fact that with quantum confinement one can achieve transition temperatures
higher than what is encountered in bulk GaAs. This may be able to 
explain the experimental observation of
spin polarized bands at room temperature found in recent spin resolved photoemission experiments\cite{arxiv-2014}.

A similar analysis was carried out for Mn doped GaN to determine the tight binding parameters as well as onsite
energies. The extracted parameters \cite{suppl-info} gave us a  $\Delta$ of 1.27 eV for Mn doping in bulk GaN and the hole has 0.516 Mn
$d$ character ($n_d$) (Fig. 4). A small decrease of $\Delta$ to 1.0 eV reduces the $n_d$ to 0.487.
This places $\Delta_{eff} = $ 0 at $\Delta \sim $ 1.15 eV. Our analysis ignored the width of the Mn $d$ states and the N $p$ states
and defined $\Delta_{eff}$ with respect to the centroid of the Mn $d$ band. Taking this into account we can place 
Mn doped GaN in the regime where $\Delta_{eff}$ is negative.
It is then surprising that we have
an insulating ground state. This phase has been called the covalent insulator
and has been shown to exist
in the ZSA phase diagram by Sarma and co-workers\cite{Nimkar-1993,DD-pramana}. Strong covalency between the transition metal and the anion are
responsible for the insulating state. It is evident from its location in the
phase diagram that any increase in $\Delta$ arising from quantum confinement
effects would not be useful in tuning the magnetic transition temperature in
Mn doped GaN.

The electronic structure of Mn doped GaAs and GaN have been examined within a multiband Hubbard model. The former may be placed
in the $p$-$d$ metal regime of the ZSA phase diagram or its variants while the latter belongs to the covalent insulating regime. 
Quantum confinement allows us to tune the effective charge transfer energy and its effect on the ferromagnetic
ordering temperature depends on where they lie within the ZSA phase diagram. This then provides us with a route to
higher ordering temperatures in the dilute magnetic semiconductors.

BM acknowledges CSIR, India for financial support.
HKC and PM thanks the Department of Science and Technology, India.

\newpage
\begin{table}
 \caption{Energy difference between q=(0.5 0.5 0.5) and q=(0 0 0) for Mn atoms separated by 11.3 \AA{} for
 1 Mn doped in 64 atoms supercell of GaAs}
 \begin{tabular}{|c|c|c|c|c|c|c|}
 \hline
 $\Delta$ (in eV) &  E[q=(0.5 0.5 0.5)]-E[(q=0 0 0)] (in eV) \\
 \hline
 0.53 &  -0.002 \\
 \hline
 1.00 &  -0.005 \\
 \hline
 1.50 &  -0.007 \\
 \hline
 2.00 &  -0.009 \\
 \hline
 2.50 &  -0.011 \\
 \hline
 2.60 &  -0.011 \\
 \hline
 2.80 &  -0.044 \\
 \hline
 2.90 &  -0.049 \\
 \hline
 3.00 &  -0.070 \\
 \hline
 \end{tabular}
\vskip2.5cm

\end{table}

\begin{table}
 \caption{Energy difference between the ferromagnetic(FM) and the antiferromagnetic(AFM) configurations
 for 2 Mn atoms occupying FCC $1^{st}$ and $4^{th}$ neighbor positions in a 64 atoms supercell of GaAs}
 \begin{tabular}{|c|c|c|}
  \hline
   & 1st nearest neighbor & 4th nearest neighbor \\
  \hline
  $\Delta$ (in eV) & E(AFM-FM) in eV& E(AFM-FM) (in eV) \\
  \hline
  1.5 & 0.120 & 0.048 \\
  \hline
  1.7 & 0.133 & 0.056 \\
  \hline
  1.8 & 0.140 & 0.057 \\
  \hline
  1.9 & 0.147 & 0.059 \\
  \hline
  2.1 & 0.161 & 0.067 \\
  \hline
  2.2 & 0.167 & 0.069 \\
  \hline
  2.3 & 0.174 & 0.072 \\
  \hline
  2.4 & 0.146 & 0.080 \\
  \hline
  2.5 & 0.116 & 0.083 \\
  \hline
  2.6 & 0.065 & 0.099 \\
  \hline
  2.9 & 0.011 & 0.074 \\
  \hline
 \end{tabular}

\end{table}

\begin{figure}[h]
\includegraphics[width=3.5in,angle=0]{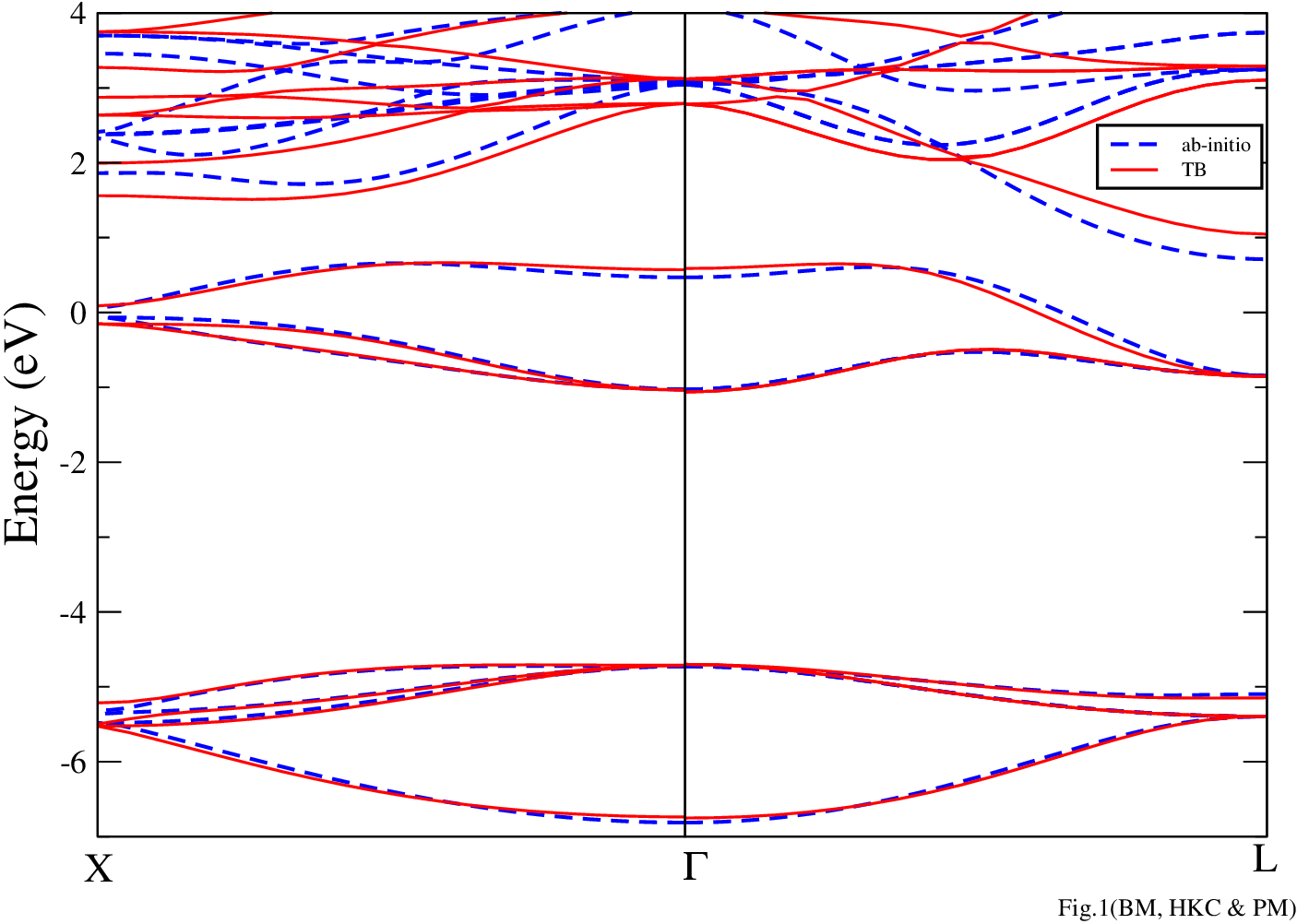}
\caption{Comparison of the ab-initio (dashed blue lines) and best fitted tight-binding (solid red lines)
band dispersions for nonmagnetic 25$\%$ Mn doped GaAs}
\vskip2.5cm
\end{figure}

\begin{figure}[h]
\includegraphics[width=3.5in,angle=0]{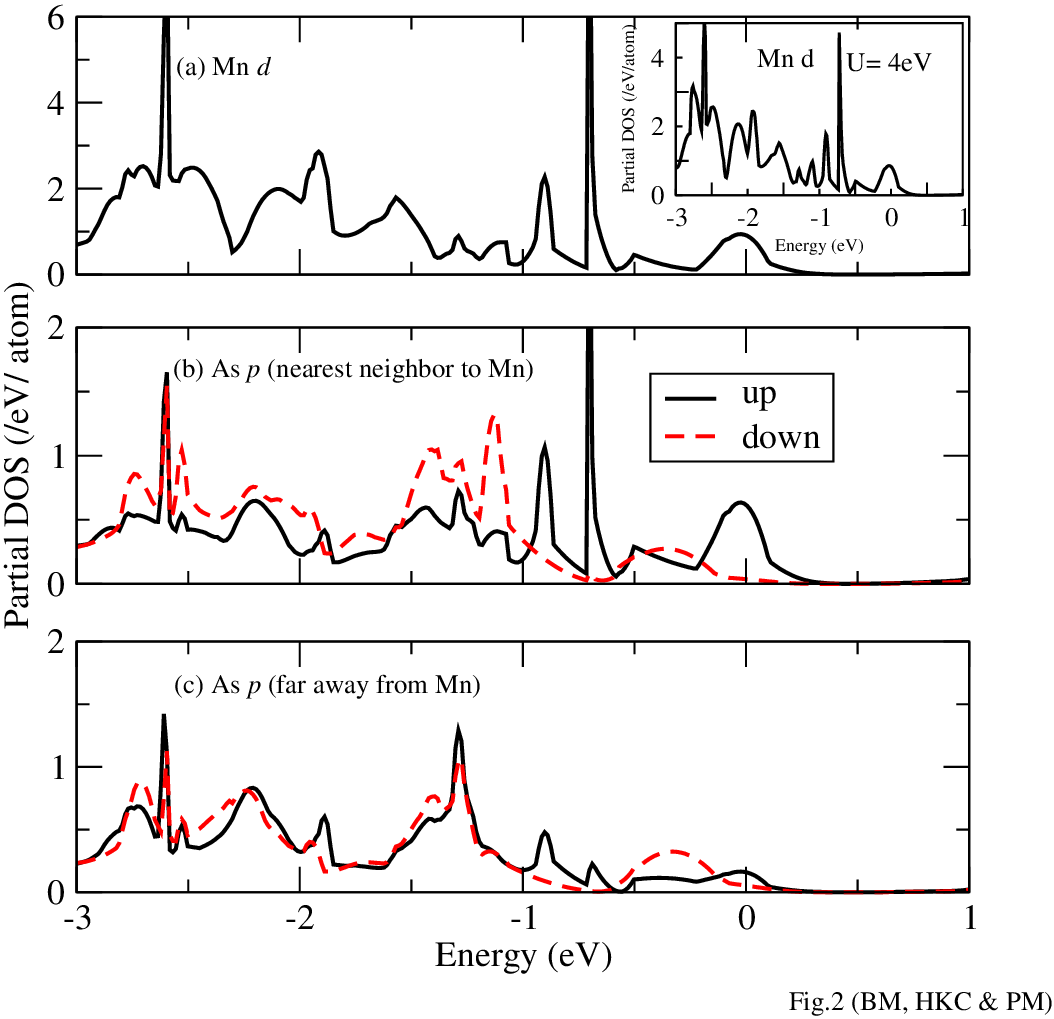}
\caption{Calculated partial density of states for (a) up spin Mn $d$, up and down spin of
(b) As $p$ (nearest neighbor of Mn) and (c) As $p$ (far away from Mn) for a Mn doping concentration of 3.125$\%$ in GaAs
at $\Delta=0.53$ eV and U=3.0 eV in a multiband Hubbard model. The zero of energy represents the Fermi energy.
Inset shows Mn $d$ density of states for up spin at U=4.0 eV.}
\vskip2.5cm
\end{figure}

\begin{figure}[h]
\includegraphics[width=3.5in,angle=0]{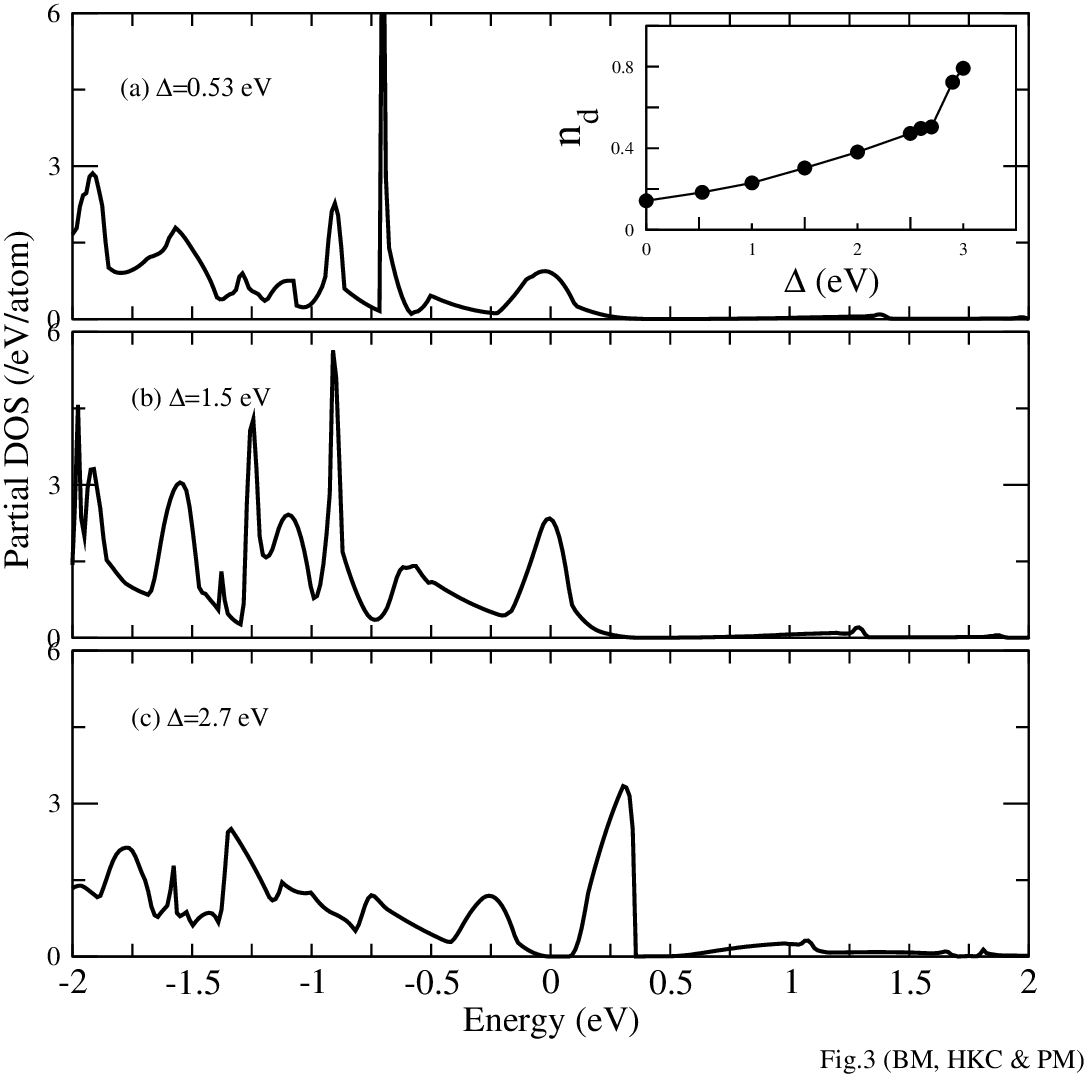}
\caption{Up spin Mn $d$ partial density of states for (a) $\Delta=0.53$ eV, (b) $\Delta=1.5$ eV and (c) $\Delta=2.7$ eV
calculated for a Mn concentration of 3.125$\%$ and a U of 3.0 eV within a multiband Hubbard model.
The zero of energy represents the Fermi energy.
Inset shows the Mn $d$ component of the hole character as $\Delta$ is varied.}
\vskip2.5cm
\end{figure}

\begin{figure}[h]
\includegraphics[width=3.5in,angle=0]{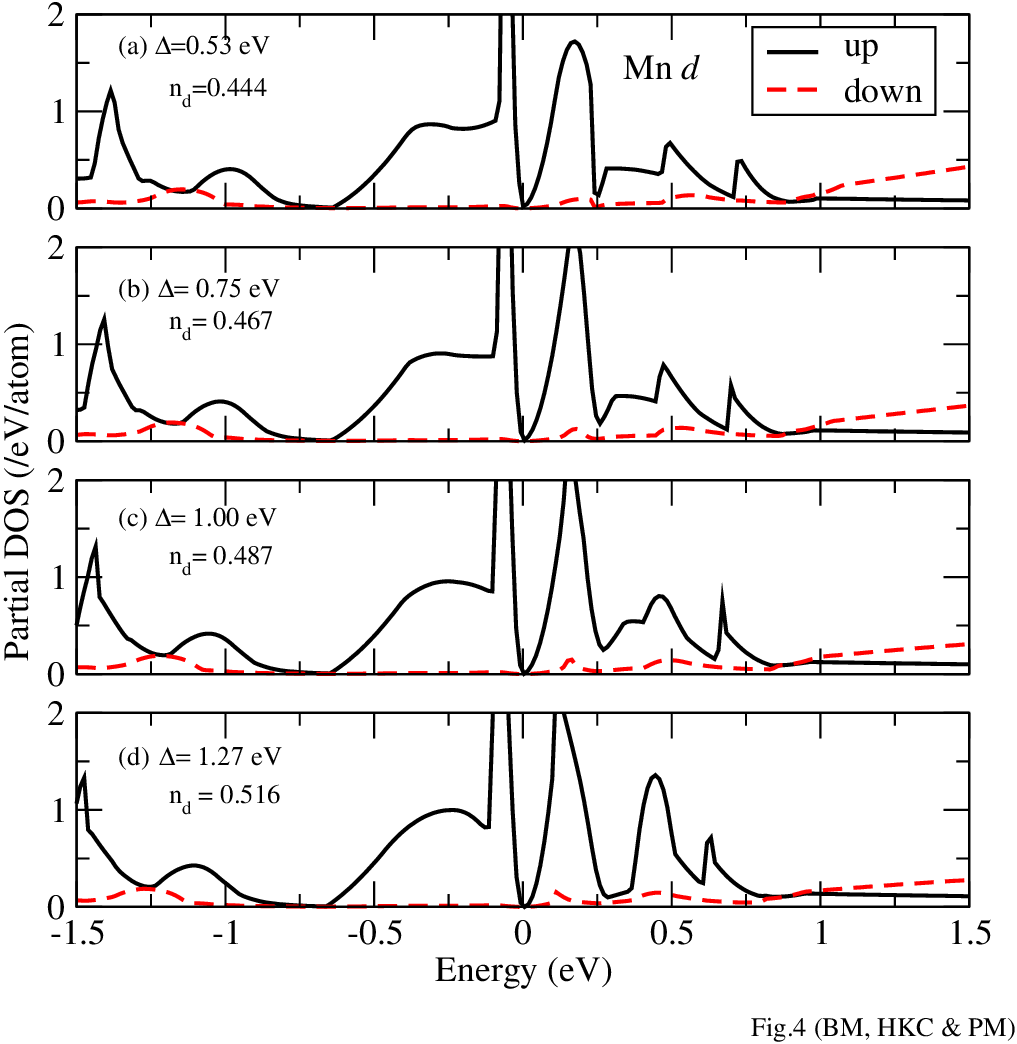}
\caption{Up and down spin Mn $d$ partial density of states for (a) $\Delta=0.53$ eV, (b) $\Delta=0.75$ eV, (c) $\Delta=1.0$ eV and
(d) $\Delta=1.27$ eV for a doping concentration of 3.125 $\%$ of Mn in GaN calculated within a multiband Hubbard model
for U=3.0 eV. The zero of energy is the Fermi energy. The Mn $d$ character ($n_d$) of the hole state has been indicated.}

\vskip2.5cm
\end{figure}


\begin{thebibliography}{0}
 
 \bibitem{Furdyna-1988}
 Semiconductors and Semimetals, Vol. 25, edited by J. K. Furdyna and J. Kossut, (Academic Pres, New York, 1988);
 C. Delerue, M. Lannoo and G. Allan, Phys. Rev. B {\bf 39}, 1669 (1989).
 \bibitem{Corbett-1983}
 J. Schneider, Defects in Semiconductors II, Symposium Proceedings, edited by S. Mahajan and J. W. Corbett,
 (North Holland, Amsterdam 1983), p.225; B. Clerjaud, J. Phys. C {\bf 18}, 3615 (1985).
 
 \bibitem{Ohno-1996}
 H. Ohno, A.Shen, F. Matsukura, A.Oiwa, A. Endo, S. Katsumoto and Y. Iye, Appl. Phys. Lett. {\bf 69}, 363 (1996).
 \bibitem{samarth-2002}
 S. J. Potashnik, K. C. Ku, R. Mahendiram, S. H. Chun, R. F. Wang, N. Samarth and P. Schiffer, Phys. Rev. B {\bf 66}, 012408 (2002).
 \bibitem{nature-2005}
 A. H. MacDonald, P. Schiffer and N. Samarth, Nat. Mater. {\bf4}, 195 (2005).
 \bibitem{rmp-2014}
 T. Dietl and H. Ohno, Rev. Mod. Phys. {\bf86}, 187 (2014).
 \bibitem{freeland-2003}
 D. J. Keaney, D. Wu, J. W. Freeland, E. Johnston-Halperin, D. D. Awschalom and J. Shi, Phys. Rev. Lett. {\bf 91}, 187203 (2003).
 \bibitem{Novak-2008}
V. Nov\'ak, K. Olejn\'{\i}k, J. Wunderlich, M. Cukr, K. V\'yborn\'y, A.W. Rushforth, K.W. Edmonds, R.P. Campion, B.L. Gallagher, Jairo Sinova, and T. Jungwirth,
Phys Rev Lett {\bf 101}, 077201 (2008).
 \bibitem{arxiv-2014}
 J. Kanski, L. Ilver, K. Karlsson, M. Leandersson, I. Ulfat and J. Sadowski, arXiv:1410.8842 (2014).
 \bibitem{APL-2001}
 S. J. Potashnik, K. C. Ku, S. H. Shun, J. J. Berry, N. Samarth and P. Shiffer, Appl. Phys. Lett. {\bf79}, 1495 (2001).
\bibitem{gamnn}
E. Sarigiannidou, F. Wilhelm, E. Monroy, R. M. Galera, E. Bellet-Amalric, A. Rogalev, 
J. Goulon, J. Cibert, and H. Mariette, Phys. Rev. B {\bf 74}, 041306 (2006).
 \bibitem{Dietl-1997}
 T. Dietl, H. Ohno, F. Matsukura, J. Cibert and D. Ferrand, Science {\bf 287}, 1019 (2000).
 \bibitem{RKKY}
 P. Mahadevan, A. Zunger and D. D. Sarma, Phys. Rev. Lett. {\bf93}, 177201 (2004).
 \bibitem{APL}
 P. Mahadevan and A. Zunger, Appl. Phys. Lett. {\bf85}, 2860 (2004).
 \bibitem{Mahadevan-2004}
 P. Mahadevan and A. Zunger, Phys. Rev. B {\bf 69}, 115211 (2004).
\bibitem{tanaka}
Shinobu Ohya, Iriya Muneta, Pham Nam Hai and Masaaki Tanaka, 
Phys. Rev. Lett {\bf 104}, 167204 (2010).

\bibitem{dobrowolska}
M. Dobrowolska, K. Tivakornsasithorn, X. Liu, J.K. Furdyna, M. Berciu, K.M. Yu
and W. Walukiewicz, Nat. Mater. {\bf 11}, 444 (2012).

\bibitem{fujii}
J. Fujii, B.R. Salles, M. Sperl, S. Ueda, M. Kobata, K. Kobayashi, 
Y. Yamashita, P. Torelli, M. Utz, C. S. Fadley, A.X. Gray, J. Braun, H. Ebert,
I. Di Marco, O. Eriksson, P. Thunstr$\ddot{o}$m, G. H. Fecher, H. Stryhanyuk,
E. Ikenaga, J. Minar, C.H. Back, G. ven der Laan and G. Panaccione, Phys. Rev.
Lett. {\bf 111}, 097201 (2013).

\bibitem{dmft}
I. Di Marco, P. Thunström, M.I. Katsnelson, J. Sadowski, K. Karlsson, 
S. Lebègue, J. Kanski and O. Eriksson, Nat. Comm. {\bf 4}, 2645 (2013).

 \bibitem{ZSA-1985}
 J. Zaanen, G. A. Sawatzky and J. W. Allen, Phys. Rev. Lett. {\bf 55}, 418 (1985).
\bibitem{Nimkar-1993}
S. Nimkar, D. D. Sarma, H. R. Krishnamurthy and S. Ramasesha, Phys. Rev. B {\bf 48}, 7355 (1993).
\bibitem{DD-pramana}
D. D. Sarma, H. R. Krishnamurthy, S. Nimkar, S. Ramasesha, P. P. Mitra and
T. V. Ramakrishnan, Pramana {\bf38}, 531 (1992).
 \bibitem{Joubert}
G. Kresse and D. Joubert, Phys. Rev. B {\bf 59}, 1758 (1999).
 \bibitem{vasp}
 G. Kresse, and J. Furthm$\ddot{u}$ller, Phys. Rev. B {\bf 54}, 11169 (1996).
\bibitem{vasp1}
G. Kresse, and J.~Furthm$\ddot{u}$ller, Comput. Mat. Sci. {\bf 6}, 15 (1996).
\bibitem{suppl-info}
See Supplementary Information for more details.
  \bibitem{PRB-TB}
 P. Mahadevan, N. Shanthi and D. D. Sarma, Phys. Rev. B {\bf 54}, 11199 (1996).
\bibitem{La2NiO4-1997}
P. Mahadevan, K. Sheshadri, D. D. Sarma, H.R. Krishnamurthy and Rahul Pandit, Phys. Rev. B {\bf 55}, 9203 (1997).
\bibitem{eschrig-1998}
S. V. Halilov, H. Eschrig, A. Y. Perlov and P. M. Oppeneer, Phys. Rev. B {\bf 58}, 293 (1998).
\bibitem{minami}
A. Fujimori and F. Minami, Phys. Rev. B {\bf 30}, 951 (1984).
\bibitem{sapra-2002}
S. Sapra, D. D. Sarma, S. Sanvito and N. A. Hill, 
Nano Lett. {\bf 2}, 605 (2002).
\bibitem{Q-confinement}
R. Viswanatha, S. Sapra, T. Saha-Dasgupta and D. D. Sarma, Phys. Rev. B {\bf72}, 045333 (2005).
\bibitem{APL-cluster}
 P. Mahadevan, J. M. Osorio-Guillen and A. Zunger, Appl. Phys. Lett. {\bf86}, 172504 (2005).
\bibitem{raebiger3}
M. van Schilfgaarde and O. N. Mryasov, Phys. Rev. B {\bf 63}, 233205 (2001).
\bibitem{kanski-prb}
I. Ulfat, J. Kanski, L. Ilver, J. Sadowski, K. Karlsson, A. Ernst and 
L. Sandratskii, Phys. Rev. B {\bf 89}, 045312 (2014).
\bibitem{akash}
A. Chakraborty, R. Bouzerar, S. Kettemann and G. Bouzerar, Phys. Rev. B {\bf85}, 01420 (2012).
\bibitem{APL-2004}
 G. Bouzerar, T. Ziman and J. Kudmovsky, Appl. Phys. Lett. {\bf85}, 4941 (2004).
 \bibitem{APL-2006}
 T. Hynninen, H. Raebiger, J. von Boehm and A. Ayuela , Appl. Phys. Lett. {\bf88}, 122501 (2006).
 
 
\end{thebibliography}
\end{document}